\documentclass[final,5p,times,twocolumn]{elsarticle}
\usepackage{amssymb}
\usepackage{graphicx}
\usepackage{dcolumn}
\usepackage{amsmath}

\begin{document}

\begin{frontmatter}

	\title{Two-band model and RVB-type states: application to Kondo lattices, pyrochlores and Mn-based systems}
\author{V. Yu. Irkhin}

\ead{Valentin.Irkhin@imp.uran.ru}
\author{Yu. N. Skryabin}
\address
{M. N. Mikheev Institute of Metal Physics, 620108 Ekaterinburg, Russia
}

\begin{abstract}
An exotic fractionalized Fermi-liquid FL$^*$ theory of metallic systems, which combines resonant-valence-bond (RVB) state and the band of current carriers,  is treated. An application of this theory to spin-liquid, antiferromagnetic  and nearly antiferromagnetic  systems is proposed  with the use of various  bosonic and fermionic representations, a comparison with perturbation theory in the $s-d(f)$ exchange model being performed. The topological aspects including formation of the small Fermi surface are treated.  In the case of narrow bands (strong correlations), the ground state  is considered as a direct product of RVB and dopon or Weng's fermion states. Examples of Kondo lattices, doped pyrochlores, and metallic $\beta$-Mn, YMn$_2$, Y${}_{1-x}$Sc${}_x$Mn${}_2$ systems are discussed, analogies with copper-oxide systems being treated. 
\end{abstract}


\end{frontmatter}

\section{Introduction}

The  resonant-valence-bond (RVB) theory was developed in the pioneering works by Anderson \cite{633b,633a}, who considered for the superconducting cuprates the concept of spin-charge separation with introducing exotic quasiparticles  -- charged bosons (holons) and neutral fermions (spinons). Early treatment concerned uniform RVB state with $T$-linear specific heat in the insulating phase. However, later more complicated approaches were elaborated, the magnetic frustrations playing an important role, see the reviews \cite{Nagaosa,Scr2}. 

Moreover, to describe all the  electron states of cuprates, a nodal--antinodal dichotomy is required \cite{Nagaosa}: the picture of the spectrum changes when passing the Fermi surface.  Near the nodal points $(\pm\pi/2,\pm\pi/2)$, the spectrum is formed by gapless Dirac quasiparticles, but has a gap at the antinodal point $(0,\pi)$. Near a nodal point (but not near the antinodal one), there occurs a  mixing between auxiliary spinons  and current carriers (dopons, according to the treatment in Ref. \cite{Ribeiro}) owing to  effective hybridization. This  picture is close in another two-band situation -- in the Kondo lattices. Here, the localized and itinerant electron states are mixed in the mean-field approach including the  $d(f)$-pseudofermions, so that the pseudofermions  take  part in the Fermi surface.

Besides cuprates, anomalous behavior occurs in metallic systems which demonstrate both the Kondo effect and magnetic frustrations, especially in anomalous (Kondo-lattice and heavy-fermion) $f$-systems. 
They can possess  reduced antiferromagnetic (AFM) moment or be near the quantum critical point -- at or near the magnetic instability  (see \cite{IK97,Si,I17,IK92}). 
As a result of competition between intersite magnetic interactions and Kondo effect (screening of magnetic moments owing to resonance scattering), and their mutual renormalization, a unique energy and temperature scale should  form in the strong-coupling regime -- the Kondo temperature $T^*_K$ \cite{IK97}. 
Unlike the crossover scenario in the case of one Kondo impurity, we can have in the Kondo lattices true phase transitions  with participation of magnetic ordering.
In the presence of magnetic frustrations, the situation becomes still more complicated since we have one more tuning parameter owing to the frustrations. 
Various scenarios of the ground state formation in such Kondo lattices  were proposed, see \cite{Si,Coleman1,Vojta}. 

A similar situation occurs also in some $d$-systems (early examples of a spin-liquid-like behavior in systems with spin and charge degrees of freedom were considered in Refs.\cite{IK90,IK93}), an ``old'' RVB theory \cite{633b} being applied. In particular,  some Mn-based compounds demonstrate giant $\gamma$T-linear specific heat or even non-Fermi-liquid behavior  \cite{Shinkoda,Stewart,Wada}. 
Usually the anomalous properties of metallic frustrated $f$-systems  are treated within the Kondo-lattice theory and of $d$-systems  within Moriya's itinerant spin-fluctuation  magnetism \cite{Lacroix}.
However, the picture of weak itinerant magnetism  is hardly  well suited in the situation with large Mn magnetic moments.

The most important and interesting effect of the frustrations is disordering of local moments, so that only partial delocalization of electron states, unlike standard Kondo lattices. Thus the system naturally described in the two-band picture including localized and itinerant degrees of freedom. The volume of the Fermi surface is determined  by conduction electrons only (a so-called small Fermi surface). In the absence of increasing the unit cell (e.g., doubling which occurs in the case of antiferromagnetic ordering), this means violation of the Luttinger theorem which states conservation on the volume enclosed by the Fermi surface with inclusion of interaction. The violation is connected with the global topological excitations associated with the emergent gauge field in the system of  deconfined spinons \cite{Sachdev,Punk1}.


A description of such anomalous frustrated metallic systems is provided  by the fractionalized Fermi liquid (FL$^*$) picture \cite{Sachdev}, which was initially formulated within the framework of the $s-d(f)$ exchange (Kondo) model. In this state, which is a kind of metallic spin liquid and has an essentially topological nature, charged excitations have conventional quantum numbers, but these coexist with additional fractionalized degrees of freedom in a second band. The  FL$^*$ was applied to frustrated Kondo lattices \cite{Vojta}.
 Also, some considerations within  the Kitaev-Kondo lattice model were performed, including realization of FL$^*$ state and  unconventional superconducting phases \cite{Balents1,Balents2}.
 

In the present paper we provide   a modern topological description  of such $d$- and $f$-compounds   within the framework  of the FL$^*$ picture and related theories, a comparison with simple perturbation theory being performed. We also treat the limit of strong correlations, where the picture of local moments is most physically clear. Although concepts of the FL$^*$ theory in various versions were formulated earlier, it seems to be timely to apply them to contemporary experimental situation in appropriate forms.

\section{FL$^*$ state: between Fermi liquid  and antiferromagnetism}

RVB  state can be identified with a quantum spin-liquid state -- long-range entangled state possessing fractionalized excitations, which is observed in a number of frustrated systems.
The theory of insulating frustrated pyrochlores was elaborated in detail in Ref. \cite{Balents} where a  classification of symmetric U(1) and Z$_2$ spin liquids on the pyrochlore lattice was performed  within the projective-symmetry-group framework for Fermi spinons, a comparison with the Schwinger boson approach being performed. 
With account of gauge fluctuations, the low temperature specific heat in the nodal-star spin liquid was calculated, the leading contribution from the bare spinons being proportional to $T^{3/2}$.

Here we focus on frustrated non-Fermi-liquid metallic systems which are subject of topological characterization, so that the FL$^*$ theory and its analogs seem to be relevant. The most important topological feature of FL$^*$ is that it possesses the small Fermi surface which does not include localized $d(f)$-electrons, unlike the usual heavy Fermi liquid (FL). 

According to the Lieb-Schulz-Mattis theorem, a  state with a gap and unbroken symmetry must have a ground-state degeneracy which has a topological nature. The existence of gapped vison excitations (in the two-dimensional (2D) case) means the existence of topologically distinct sectors on a torus and  protects the FL$^*$ state.
The combination of the spin-liquid state and current carriers can also lead to violations of the area law for the entanglement entropy \cite{Vojta,Vojta1}.


Usually the FL$^*$ concept is applied to 2D case, but can be generalized to 3D case owing to Oshikawa’s theorem \cite{Oshikawa,Sachdev}.
The FL$^*$ state was considered near both the metal-insulator transition  and  transition between metallic states with large and small Fermi surfaces,  and includes fermionic spinon excitations which form a ``ghost'' Fermi surface. The corresponding phase diagrams were built in both doped case and the situation of interaction-driven transition \cite{Sachdev,Senthil1,Senthil2}.

In the gauge field theory, low-temperature specific heat has a singularity  $C(T) \sim T \ln 1/T$  in the quantum-critical region and U(1) FL$^*$ phase or $ C(T) \sim T \ln 1/b_0$ in the Fermi liquid near the transition, $b_0$ being the auxiliary-boson condensate determining the mixing between local moments and conduction electrons
at FL formation. 
Thus the specific heat coefficient $\gamma = C/T$ diverges logarithmically in the FL$^*$ phase (in the 2D case, $C(T) \sim T^{2/3}$)  \cite{Sachdev,Senthil2}. 
Besides that, $T$-linear resistivity occurs in the quantum-critical strange metal phase \cite{Senthil1}.

The FL$^*$ state was considered as a ground
state \cite{Sachdev,Senthil1,Senthil2}. However,  it was also concluded  \cite{Sachdev,Sachdev1}
that this state with the spinon Fermi surface should be unstable with to a  broken-symmetry AFM state, but can occur at finite temperatures.
With increasing $s-d(f)$ coupling, the deconfined phase with small Fermi surface passes first into usual itinerant AFM state with a large  Fermi surface volume, and then into FL state.
According to \cite{Vojta}, a direct transition from localized AFM state to the Kondo FL state in $T=0$  phase diagram is possible only  with fine tuning via the multicritical point. 
Some different versions of the phase diagram were treated \cite{Si,Coleman1,Punk1}. In particular, the approach \cite{Si} yields a whole AFM-FL line. This situation  may be viewed as an example of deconfined criticality \cite{Vojta}.



Most Kondo and heavy-fermion systems demonstrate magnetic ordering and/or pronounced spin fluctuations even in the strong-coupling regime \cite{I17}.
A transition (crossover) from itinerant to localized magnetism  in the Kondo lattices is to some extent analogous to the metal-insulator transition (localized moments correspond to the Hubbard subbands). Moreover, the transition to FL$^*$ phase can be treated as a partial (orbital-selective) Mott transition \cite{Vojta}.

Provided that the quantum critical point and magnetic transition point coincide, the FL$^*$ theory includes two different divergent time or length scales \cite{Senthil2}. The first (shorter) scale describes fluctuations owing to the  Fermi surface rearrangement, and the second one due to magnetic  fluctuations. 
Thus, on the magnetically-ordered side of the
quantum phase transition into the FL state, there should
be an intermediate temperature regime below the coherence temperature, $T_N < T < T^*_{coh}$. where we have the FL$^*$ picture \cite{Sachdev}. 
Such a picture was proposed to treat  properties of the frustrated layered system PdCrO$_2$ \cite{Komleva}. 
Near the metal-insulator transition, a second crossover (which is connected with condensation of spinons), and even some more crossovers, were proposed \cite{Senthil1}.
Such crossovers can be supposed also on both above magnetic and FL phases of the Kondo lattice.



The description in terms of spinon statistics  is changed from fermionic to bosonic in AFM phase after passing deconfinement transition, so that the description in terms of auxiliary particles becomes different \cite{Punk1,Ramires}. In particular, the Schwinger bosons are convenient to describe AFM state, and the pseudofermions to describe the Fermi-spinon  picture.
In this connection, a supersymmetry approach should be mentioned \cite{Ramires}, which uses both Fermi ($f$) and Bose ($b$) operators with corresponding constraints,
\begin{equation}
	\mathbf{S}_{i}   = (1/2)\sum_{\alpha \beta }\mbox {\boldmath $\sigma $}_{\alpha \beta }
	(f_{	i\alpha }^{\dagger }f_{		i\beta }^{}+b_{	i\alpha }^{\dagger }b_{		i\beta }^{}).
	\label{H11}
\end{equation}
with $\mathbf{\sigma }$  the Pauli matrices.

In the AFM localized-moment phase (AFM FL according to \cite{Punk1}), the long-range order can be described as  condensation of Schwinger bosons \cite{IK91}. With increasing $T$ above the (low) Neel temperature $T_N$  both in  quasi-2D \cite{IK91} and frustrated 3D case \cite{IKK}, the condensate disappears, but the correlation length remains large, so that we have a quasi-condensate situation.
Then a strong short-range order and spin-wave picture of spectrum in the Heisenberg antiferromagnet hold in a  broad range above $T_N$ (in the renormalized classical regime).
In such a situation, interaction with magnons results in a splitting of electron spectrum \cite{IK91} and in a marginal behavior of electron properties in some temperature interval \cite{IK95}. 
Therefore we can perform comparison with simple perturbation theory in the standard broad-band $s-d(f)$ model.
This theory does not provide description of the ground state, but may give some indications of violating the FL behavior starting from high temperatures  \cite{IK95,pert}.

The corresponding Hamiltonian  reads
\begin{equation}
	H=\sum_{ij\sigma }t_{ij}c_{i\sigma }^{\dagger }c_{j\sigma }-I\sum_{i\alpha \beta }(\mathbf{S}_i \cdot 
	\mbox {\boldmath $\sigma $}_{\alpha \beta })
	c_{i\alpha }^{\dagger }c_{%
		j\beta }+\sum_{ij}J_{ij}(\mathbf{S}_{i}\cdot%
	\mathbf{S}_{j})  \label{H}.
\end{equation}
Here $c_{i\sigma }^{\dagger }$, $c_{j\sigma }$ and $\mathbf{S}_{i}$ are operators for conduction electrons and localized moments, $I$ is the $s-d(f)$ exchange
parameter. 

The part of  electronic specific heat owing to interaction with localized spins is obtained by using the Hellmann-Feynman theorem for the free energy, 
\begin{equation}
\partial 
\mathcal{F}/\partial I=\langle \partial H/\partial I\rangle = 
-\langle 
(\mathbf{S}_i \cdot 
\mbox {\boldmath $\sigma $}_{\alpha \beta })
c_{i\alpha }^{\dagger }c_{	j\beta}\rangle,  
  \label{H123}
\end{equation}
and calculating the correlation function to second order in $I$.
 We treat the  most important contribution  from transitions between AFM subbands. They work, owing to 
conservation laws, at $|\mathbf{q-Q}|>q_0\sim
\Delta /v_F$ with $\bf Q$ being the wavevector of the AFM structure, $\Delta =2|I\overline{S}|$  the AFM splitting of electron band, $\overline{S}$  the  staggered moment, $v_F$ the average electron
velocity at the Fermi surface. We obtain (see  \cite{pert})
\begin{equation}
	\delta C_{}(T)=\frac 4 3\pi^2\Delta^2(J_0-J_{\mathbf{Q}})
	\sum_{\mathbf{k},\mathbf{q\simeq Q,}\omega _{\mathbf{q}}\leq T}
\delta (t_{\mathbf{k}})\delta (t_{\mathbf{k+q}})/\omega _{\mathbf{q}}^2  \label{intc}
\end{equation}  
where the electron spectrum $t_{\mathbf{k}}$ is counted from the Fermi level, $\omega _{\mathbf{q}}$ is the usual magnon frequency in the zeroth order in $I$ (i.e., for the Heisenberg Hamiltonian). In the $2D$ or quasi-$2D$ case the integral is logarithmically divergent at $\mathbf{q}\rightarrow {\bf Q}$, the divergence being cut at  $\omega _{\mathbf{q}}\simeq \max (T,T^{*})$ where
\begin{equation}
	T^{*}=cq_0\sim J\overline{S}^2\Delta /v_F \ll J\overline{S}^2 \label{T*},
\end{equation}
with $c$ being the magnon velocity for $\mathbf{q}\rightarrow {\bf Q}$, 
 $\omega _{\mathbf{q-Q}}=cq$. Picking out the singular contribution we obtain 
\begin{equation}
	\delta C_{}(T)\sim
	T	\ln \frac{\bar \omega }{\max (T,T^{*})}  \label{cinter}.
\end{equation}
The integral over $\mathbf{q}$ in (\ref{intc}) is determined by the magnon spectrum. Therefore, the result (\ref{cinter}) is valid also in the 3D frustration situation, e.g., for the $2D$-like magnon spectrum 
\begin{equation}
	\omega _{\mathbf{q-Q}}^2=c_x^2q_x^2+c_y^2q_y^2+c_z^2q_z^2+	...
\end{equation} 
with $c_z\ll c_x,c_y$ (another frustration model is considered in Ref. \cite{IK92}). Thus at $T>T^{*}$ we have a marginal Fermi liquid behavior: the $T\ln T$-dependence of specific heat, and also $T\ln T$ (nearly $T$-linear) dependence of resistivity \cite{pert}.
Therefore, we have an interpolation with the results of the FL$^*$ theory.

To take into account magnetic fluctuations above the FL state (at the crossover in the strong-coupling region),  we can treat fluctuations of the effective $s-d(f)$ hybridization  which are most important in the Kondo lattices. The standard representation of the scalar slave boson (measuring the  hybridization)
is insufficient to take into account spin-flip processes. Therefore we can use at finite $T$ (in the renormalized classical regime \cite{Sachdev}) the representation  \cite{Isaev} in terms of pseudofermions $f_{\alpha}$ and bosons $b_{\mu}$,
\begin{equation}
	b_{i\mu}=\sigma^\mu_{\alpha\beta} f^\dag _{i\alpha}c_{i\beta}/\sqrt{2}
	\label{is}
\end{equation}
with $\mu=0\ldots3$ and $\sigma^0_{\alpha\beta}=\delta_{\alpha\beta}$.
The operators $b_{i\mu}$ are  Schwinger bosons which create local singlet and triplet states resulting from the $s-d(f)$ coupling between localized and itinerant states. In the usual picture of Kondo singlet formation, only $b_0$-boson is present, and other components describe the magnetic order fluctuations. The scattering by such fluctuations should result in a similar behavior of thermodynamic and transport properties, as described above.


To describe the exotic state with small Fermi surface in the strong-interaction case (narrow bands), the approaches by Weng (the phase string theory \cite{Weng,Weng1}) and Ribeiro-Wen \cite{Ribeiro} were proposed.  These formulations differ by statistics of auxiliary particles but seem to be on the whole equivalent. Both the approaches include  subsystems of RVB states and conduction electrons, so that the wavefunction of the system can be represented as a direct product of half-filled fermionic or bosonic RVB state and conduction-electron dopon or Weng's fermion state, respectively  \cite{WW}. The approach using ``backflow'' fermionic spinons \cite{Weng1} should be also mentioned.

Consider the Hamiltonian of the $t-J$ model
 \begin{equation}
	H=\sum_{ij\sigma
	}t_{ij}\tilde{c}_{i\sigma  }^{\dagger
	}\tilde{c}_{j\sigma}+\sum_{ij}J_{ij}(\mathbf{S}_{i}\cdot%
\mathbf{S}_{j}) 
	\label{eq:I.51}
\end{equation}
where
\begin{equation} 
	\tilde{c}^{\dagger}_{i\sigma}=X_i(0,\sigma)=|i0 \rangle \langle i \sigma|
\end{equation} 
are  the projected electron (Hubbard's) operators.  This model is equivalent to the narrow-band $s-d$ model \cite{Scr}. The elimination of the doubly occupied states (i.e., of the upper Hubbard subband) owing to the projection just means formation of the small Fermi surface.

The dopon representation \cite{Ribeiro} for the Hubbard's operators  reads
\begin{equation}
	\tilde{c}_{i-\sigma }=-\frac{\sigma }{\sqrt{2}}\sum_{\sigma ^{\prime }}%
	\tilde{d}_{i\sigma ^{\prime }}^{\dagger }[S\delta _{\sigma \sigma ^{\prime
	}}-(\mathbf{S}_{i}\cdot\mbox{\boldmath$\sigma $}_{\sigma ^{\prime }\sigma })].
	\label{eq:I.78}
\end{equation}%
where $\sigma =\uparrow ,\,\downarrow $ ($\pm $),  $\tilde{d}_{i\sigma }^{\dagger }=d_{i\sigma
}^{\dagger }(1-d_{i-\sigma }^{\dagger }d_{i-\sigma })$ with $d_{i\sigma
}^{\dagger }$  the Fermi dopon operators. Depending on physical picture, both Fermi spinon and Schwinger boson
representations can be used for localized $S=1/2$ spins $\mathbf{S}_{i}$
\cite{Ribeiro,Punk2}. In the Bose case, the small Fermi surface is retained, but  in the Fermi case the picture can become more complicated owing to spinon-dopon hybridization. The corresponding  mean-field calculations \cite{Ribeiro} enable one to describe qualitatively the spectrum of copper-oxide systems, but a more accurate treatment including gauge field fluctuations is not yet performed. 

 Using the many-electron Hubbard's representation, the dopon representation is generalized to the narrow-band $s-d$ exchange model \cite{Scr} with arbitrary spin $S$, which is appropriate for localized-moment Mn systems:
  \begin{equation}
 	H=\sum_{ij\sigma
 	}t_{ij}g_{i\sigma  }^{\dagger
 	}g_{j\sigma}+\sum_{ij}J_{ij}(\mathbf{S}_{i}\cdot%
 \mathbf{S}_{j}),
 	\label{eq:I.5}
 \end{equation}
 where
 \begin{equation}
 	g_{i\sigma }^{\dagger }=\sum_{\sigma ^{\prime }}c_{i\sigma
 		^{\prime }}^{\dagger }(1-n_{i,-\sigma ^{\prime }})
 	\frac {S \delta _{\sigma \sigma ^{\prime }}-(\mathbf{S}_i\cdot\mbox{\boldmath$\sigma $}_{\sigma ^{\prime }\sigma })} {2S+1}.
 	\label{eq:I.8}
 \end{equation}
The result (\ref{eq:I.8}) corresponds to the double-exchange theory by Kubo and Ohata  \cite{Kubo}, which is also appropriate for manganites.


Thus in  the  cases of both strong and  weak $s-d(f)$ interactions  we have a description in terms of the two-band picture within different models.

\section{Discussion and conclusions}

Now we discuss frustration effects in real metallic systems. A number of Kondo and heavy-fermion compounds demonstrate frustrated magnetism and non-Fermi-liquid behavior (see, e.g., \cite{I17}). 
In  the distorted kagome  system CePdAl
 magnetic frustration of 4f-moments results in  a paramagnetic quantum-critical phase with non-Fermi-liquid behavior of resistivity \cite{Zhao1,Zhao}.  This picture includes also a Mott-type f-electron partial delocalization (transition from small to large Fermi surface) \cite{Zhao}.
Two-dimensional fluctuations at the quantum-critical point are observed in CePdAl  \cite{Zhao1},
CeCu$_{6-x}$Au$_{x}$ \cite{2D}
and YbRh$_{2}$Si$_{2}$ \cite{YbRh2Si2}. 

The electronic state of Y$_2$Ir$_2$O$_7$ pyrochlores is insulating. In the hole-doped Y$_{2-x-y}$Cu$_x$Ca$_y$Ir$_2$O$_7$  system, the resistivity becomes much smaller than that of pure compound and has a metallic $T$-linear behavior with a small upturn below  around 15 K.  According to  $\mu$SR measurements,  the AFM long-range ordered state occurs in pure case, while the ground state of the doped system is non-magnetic up to 2 K \cite{Angel}. Thus the antiferromagnetism is suppressed by doping,  the system  being changed from insulator to metal. From these results, the authors propose that there is a quantum critical point governed by doping.
This picture is similar to the phase diagram of copper-oxide systems 
\cite{Nagaosa}.

It is important that, upon doping, Z$_2$ spin liquid can pass to superconducting state owing to spinon pairing \cite{Sachdev}, as it is discussed in the case of cuprates in Ref. \cite{Nagaosa}. Indeed, superconductivity with rather high $T_c$ is observed in the metallic pyrochlores  KOs$_2$O$_6$ ($T_c$ = 9.6 K \cite{KOs2O6}),  RbOs$_2$O$_6$ ($T_c$ = 6.4 K \cite{RbOs2O6}, $\gamma$ = 34 mJ/mol${}\cdot {}$K${}^2$), and CsOs$_2$O$_6$ ($T_c$ = 3.3 K \cite{CsOs2O6}).

In pure $\beta$-Mn, $\gamma$ = 80 mJ/mol${}\cdot {}$K${}^2$ \cite{Shinkoda} and NFL behavior is observed \cite{Stewart};  $\gamma$ decreases upon doping  with occurrence of AFM ordering and increasing the Neel temperature \cite{Miyakawa}. 

According to Ref.\cite{Nakamura}, the magnetism
of $\beta$-Mn can be reduced to that of an infinite network of corner-sharing triangles, which is in some respects identical to the 2D kagome lattice; besides that, $\beta$-Mn is not a ``weak'' magnet, but  a rather strong magnet with a large dynamical moment up to lowest $T$. Basing on polarized-neutron scattering experiments, the authors of Ref.\cite{Paddison} demonstrate the presence of emergent spin structures resembling the  triangular-lattice antiferromagnet and collective states within  networks in Co-doped $\beta$-Mn.

According to Ref. \cite{Mn2FeAl}, in the Mn$_2$FeAl compound with the $\beta$-Mn structure, frustrated antiferromagnetism coexists with large $\gamma$ = 210 mJ/mol${}\cdot {}$K${}^2$.

The compound YMn${}_2 $   has a lattice structure similar to pyrochlores and a frustrated AFM structure, demonstrating strong short-range order above $T_{\text{N}}$. In the doped system Y${}_{1-x}$Sc${}_x$Mn${}_2$ with $x=0.03$ or under pressure,  the long-range order is destroyed, while $\gamma$  reaches
very large value, $140$ mJ/mol${}\cdot {}$K${}^2$  \cite{Wada}. 

Inelastic neutron scattering measurements on a single crystal of Y${}_{1-x}$Sc${}_x$Mn${}_2$ \cite{Ballou} demonstrate that the dynamical susceptibility indicates a degeneracy of magnetic states.
Moreover, spins seem to be correlated and form short-lived four-site collective spin singlets. 
Such a frustrated heavy-fermion behavior is quite unusual within the picture of itinerant magnetism \cite{Ballou}.

To conclude, the topological treatment within the two-band model clarifies the physical picture and  enables one to unify frustrations, Kondo and spin-fluctuation (e.g., electron-magnon) mechanisms 
which were considered earlier separately \cite{IK90,IK93}. A comparison with perturbation theory and strong correlation limit permits to obtain a consistent picture of peculiar magnetic, thermodynamic and transport properties of $d$- and $f$-compounds, especially anomalies of specific heat.

The topological picture is supported by formation of the small Fermi surface in magnetically frustrated systems (contrary to usual Kondo lattices). More generally, these arguments can be related to the problem of Hubbard subbands \cite{Scr4} and  disordered local moments which violate the Fermi liquid picture and Luttinger theorem too \cite{Auslender}. To treat these questions in more detail, a consistent consideration of gauge field effects with proper account of constraints would be instructive. The involved problem of describing spinon statistics is also far from final solution.

The research was carried out within the state assignment of the Ministry of Science and Higher Education of the Russian Federation (theme ``Flux'' No AAAA-A18-118020190112-8 and theme ``Quantum'' No. AAAA-A18-118020190095-4). 
{}
\end{document}